# The Problem of a Perfect Lens made of a Slab with Negative Refraction

A. L. Efros
*Dept. of Physics, University of Utah, UT 84112-0830, USA*

1. Introduction

Recently there has been growing interest in the creation of lenses with unusually sharp foci. Those lenses come in two types.

  **A. Veselago lens.** Moscow physicist Victor Veselago[1] originally proposed it in 1967. Veselago studied a medium where due to some reason both electric permittivity $\varepsilon$ and magnetic permeability $\mu$ are negative at some frequency $\omega_0$. Since the product $\varepsilon\mu$ is positive the light velocity is real, and the wave equation has a regular form. However in a plane wave the vectors **k, E, H** form a left-handed rather than a right-handed set, so that the Poynting vector **S** is opposite to the vector **k**. Such a medium is often called left-handed medium (LHM). This anomaly is not forbidden by any general principle. Veselago predicted negative refraction (minus sign in Snell's law) at the interface of the LHM and a regular medium (RM) and some other interesting manifestations of the LHM.

  It is important to understand that the negative refraction is a property of the interface that follows from the regular boundary conditions. We think one should be cautious when prescribing a negative value to the bulk refractive index n to the LHM[2,3] even though this definition would restore positive sign in Snell's law. The point is that neither Maxwell's equations nor the boundary conditions contain n. The algorithms required for the solution do not include the operation of taking the square root of $\varepsilon\mu$. Therefore, we do not think the sign of n has a physical meaning at all. One can define n as a negative branch of the square root for either LHM or RM but then habitual equations, like $\omega = ck/n$ and expression for a group velocity, should be changed also and this may lead to some mistakes (See details in Ref.[4]). In this paper n is considered as a positive number.

  Veselago proposed a lens, based upon negative refraction (See Fig. 1). The lens consists of a long slab of a material with negative $\varepsilon$ and $\mu$ imbedded into a RM with positive $\varepsilon$ and $\mu$ that have the same absolute values. Since impedances are matched there are no reflected waves.

  The ideas of Veselago were completely forgotten but they got a new life after J. Pendry published in 2000 a paper[5] with a startling title "Negative refraction make a perfect lens".

    **B. Electrostatic or quasistatic lens.** Research on this lens goes back to the paper by Nicorovici *et al.*[6] of 1994 but the plane lens proposed by Pendry[5]. Consider a metallic slab and an electromagnetic field with such a low frequency

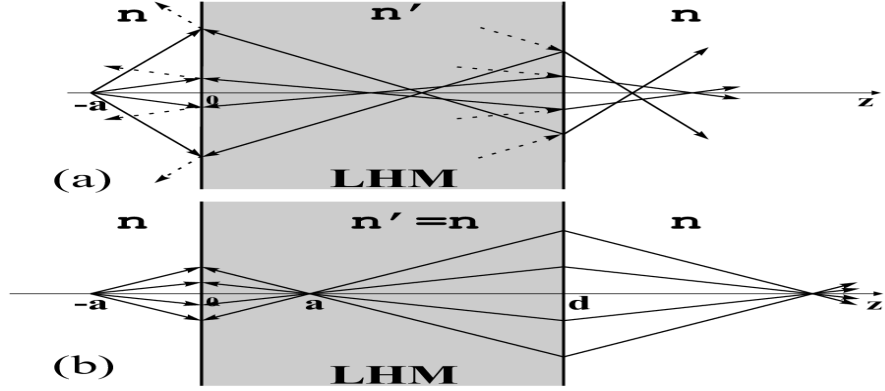

**Figure 1.** Veselago lens. The source is at point -a. The width of the LHM slab is d. The arrows of rays show direction of **k**. (a) If refractive indexes and impedances of the LHM and the RM are not matched, there is no focusing. (b) If they are matched, the lens has two foci, one at z= a, and another at z=2d-a. It is assumed that d>a.

that the wavelength is much larger than the width of the slab. Then one can neglect retardation. In this approximation magnetic field is small and the electrostatic potential $\phi(r)$ obeys the Laplace equation

$$\nabla(\varepsilon(r)\nabla\phi(r)) = 0. \qquad (1)$$

On the other hand, if the frequency obeys condition ωτ>>1, where τ is relaxation time of electrons, the dielectric constant of metallic slab may be written in a form

$$\varepsilon(\omega) = 1 - (\omega_p/\omega)^2, \qquad (2)$$

where $\omega_p$ is the plasma frequency. To get $\varepsilon$ =-1 one should have $\omega = \omega_p/\sqrt{2}$.

All these conditions may be fulfilled only if the slab is thin enough (20-40nm). So the new physics of this approach is closely connected to the recent development of nanotechnology and rarely can be found in classical textbooks. The force lines of electric field reproduce Fig. 1 if the point charge is at z= –a. So the electrostatic slab with ε=-1 works in the same way as the Veselago lens. This was discovered by Nicorovici et al.[6] for the cylindrical lens and by Pendry[5] for the plane lens. Note, however, that the wavelength criterion for the width of the focus, which should be used for the Veselago lens, is irrelevant for the electrostatic lens because in the electrostatic approximation the wavelength is infinite.

Pendry's result for both lenses is that the fields near the foci are exactly the same as near the source (the perfect imaging). According to Pendry a small absorption inside the slab may only slightly smear the foci.

The ideas behind his paper are very straightforward. In the case of the Veselago lens they are as follows. If the slab of the Veselago lens is infinite, the lens has an infinite aperture. A source produces both propagating and evanescent waves (EW's). The latter waves exist usually in the near field region only. In the regular lens these waves would never reach the focus, and that is why the image is not ideal. However, Pendry has shown that EW's are amplified by the LHM. Moreover, he claimed that this amplification leads to a complete restoration of the EW's in a focal point. Therefore, the image should perfectly repeat the source.

In the case of electrostatic lens all waves are evanescent. Pendry claims that they also decay in the regular material and get amplified by the metallic slab up to the ideal restoration of the image in the foci.

Pendry demonstrated the amplification of a single EW. The idea looks good nevertheless it is incorrect. The perfect lens with a real focus is forbidden by Electrodynamics. It can be easily seen from the following argument. The field near the point source at z=-a behaves as 1/r, where r is a distance from the source. The field near the focus cannot behave this way. Such a behavior would not obey Maxwell's equation because there is no source near the focus. The situation with a perfect lens in the electrostatic regime is even worse. Not only is a 1/r singularity forbidden without the source but also any potential with a maximum or minimum cannot be a solution of the Laplace equation.

The flaw in Pendry's arguments was found immediately at least by three groups[7-9]: the solution in coordinate space, summarized over all EW's diverges exponentially in a 3-dimensional domain near the rear face of the slab. Thus, this is not a solution of Maxwell's equations. Note that Nicorovici *et al* [6] understood this problem in 1994.

There were numerous attempts to make regularization of this solution using small imaginary parts of $\varepsilon$ and $\mu$[10,11] or by considering time-dependent instead of stationary solution[12,13]. However, the divergences of the fields are too strong to be removed by a small regularization preserving the physical picture. Recently Milton *et al*. have shown[11,14] that if a<d/2 (see Fig. 1) the absorption of energy in the slab $\text{Im}\,\varepsilon\,|E|^2$ *diverges* as $\text{Im}\,\varepsilon$ *vanishes*. This result is valid for both types of lenses. It reflects a strong divergence of the electric field as $\text{Im}\,\varepsilon$ vanishes. Note that the increase of the fields in the case of non-zero $\text{Im}\,\varepsilon$ occurs near both faces of the slab due to reflection of the wave.

Both the power that is going through the slab and the power that is dissipated inside are provided by the source. However, real sources have a finite power. Milton *et al*.[14] argued that at a<d/2 all the energy provided by a source is dissipated in a slab. Then in the limit of small $\text{Im}\,\varepsilon$ the slab turns into something opposite to a perfect lens: it cloaks the source, makes it invisible behind the slab.

In the case d/2<a<d the divergences of the fields are weaker so that absorption tends to zero as $\text{Im}\,\varepsilon$ tends to zero. Nevertheless, the limit of the solution at $\text{Im}\,\varepsilon \to 0$ is not a solution of Maxwell equation, and it looks strange for us.

**2. Existence of the perfect lens with virtual focus.**

Now I present some modification of the above geometry such that Pendry's arguments on the amplification of EW's becomes perfectly right so that both types of the lenses under discussion are indeed the perfect lenses in some sense, and solution of Maxwell's equations exists even without absorption.

Consider a lens with a width d and an object at a distance $d_1$ from a lens as shown in Fig. 2. Suppose that $d_1>d$. Under this condition the lens does not have any real focus but Fig. 2 shows that the lens has a virtual focus at $z=2d$. In the case $d<d_1<2d$ this focus is within the slab. If $2d<d_1$ (this case is shown in Fig. 2), the virtual focus is to the left of the slab.

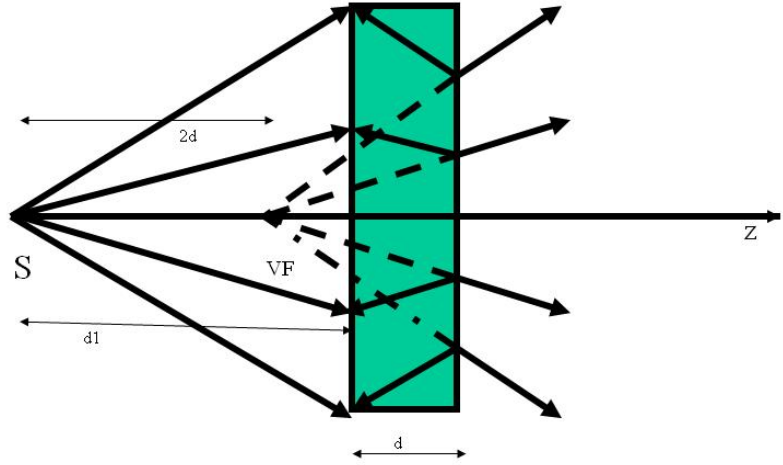

**Figure 2.** Veselago lens or electrostatic lens with the object at $z=0$ at a distance $d_1>d$ from the slab. One can see that the lens has a virtual focus (VF) at $z=2d$. It may be inside or to the left of the slab. It is important that the field in the VF does not have a singularity. An observer behind the lens sees the image located not at point S ($z=0$), where the object is, but at $z=2d$. Thus the image looks shifted at a distance $2d$ toward the observer. The lens is perfect since the image is only shifted but it is not distorted.

It is important that because fields do not have any singularity near the virtual focus there are no arguments forbidding perfect virtual focus. We show below that an observer behind the slab sees the image translated without any distortion at a distance $2d$ toward the observer. Pendry's amplification of the EW's plays a crucial role in this derivation. We consider the case of electrostatic lens but the same consideration can be done for the Veselago lens.

Following Pendry we suppose that electrostatic potential at point S ($z=0$) has a form

$$\phi(x,y,0) = \frac{1}{(2\pi)^2} \iint dk_x dk_y V(k_x,k_y) Exp(ik_x x + ik_y y), \qquad (3)$$

where time exponent is omitted. For $z>0$, the potential has a form

$$\phi(x,y,z) = \frac{1}{(2\pi)^2} \iint dk_x dk_y V(k_x,k_y) Exp(ik_x x + ik_y y) F(z), \qquad (4)$$

where the function F(z) should be found in three different regions using the equation

$(\frac{d^2}{dz^2} - k_0^2)F = 0$ with $k_0^2 = k_x^2 + k_y^2$ and boundary conditions at the faces of the slab. Assuming that ε=1 outside the slab and ε=-1 inside, one gets that F(z) is continuous while $dF/dz$ should change sign at the boundaries of the regions. Then we get

$$F(z) = Exp[(-k_0 z)] \text{ at } 0<z<d_1, \qquad (5)$$

$$F(z) = Exp[k_0(z - 2d_1)] \text{ at } d_1<z< d_1+d, \qquad (6)$$

$$F(z) = Exp[-k_0(z - 2d)] \text{ at } z>d_1+d. \qquad (7)$$

Here $k_0 = \sqrt{k_x^2 + k_y^2} > 0$. One can see a single virtual focus. An observer at any point behind the lens (z> $d_1$+d) sees potential $\phi(x, y, z)$ as if it were created by the object, given by Eq. (3) but this object is shifted without any distortion at point z=2d. The absence of a distortion is due to *amplification* of potential with increasing z inside the slab, as predicted by Pendry. Note, that all integrals over $k_x,k_y$ are finite at $d_1$>d because z-2$d_1$ <0 inside the slab and z-2d>0 behind the slab. Thus, in this case the solution of Maxwell's equation can be obtained without any regularization.

It is interesting that the virtual focus may be located infinitesimally close to the rear face of the slab (point z=d+$d_1$). It happens if the object is at a distance $d_1$ from the lens and $d_1$ is only infinitesimally larger than the thickness of the slab d. Let $d_1$=d+η, where $\eta \to 0$. Now the virtual image is inside the slab at a distance η from its rear face. At negative η the focus becomes real and the theory is ruined by divergent integrals as it happens with Pendry's theory. But at any positive η everything is fine. This means that the image without any distortions can be created infinitesimally close to the rare surface of the slab but still inside the slab.

Finally we discuss how the spatial dispersion of the dielectric constant changes our results. In the case of the Veselago lens in a photonic crystal when the working frequency is close to the Γ-point of the second Brillouin zone the propagating waves are characterized by negative $\varepsilon$, $\mu$ and they have negative refraction at the interface with a RM. But the EW's have different (**k**-dependent) $\varepsilon$ and $\mu$ due to the spatial dispersion effect so that the amplification may just be absent [15,16]. Then there are no divergences without absorption for all real and virtual foci. All of them are not perfect and to find fields in the foci one should just subtract evanescent waves from the fields of the source. Note that the amplification of the EW's in photonic crystals may be possible not due to the surface plasma modes, that result from negative $\varepsilon$ and $\mu$, but because of the surface modes that appear just due to a termination of the crystal. The existence and properties of such modes depend on a way the surface is cut. These modes may provide a superlensing in the near field regime[15].

In the case of electrostatic lens based upon metallic slab the effect of spatial dispersion cuts all wave vectors above $\omega/v_F$, where $v_F$ is the velocity of electrons at the Fermi surface. Due to this effect maximum resolution of the electrostatic lens should be about 10 nm. (See Ref.[17] for details).

3. **Conclusions**

The problem of the principal existence of the perfect lens and superlensing is discussed. We have demonstrated that in the case of the virtual focus the idea of perfect lens based upon amplification of evanescent waves as proposed by Pendry is perfectly right unlike the case of the real foci. We think that some experimental results claiming superlensing can be explained in terms of the proposed theory for the case when the virtual focus is inside the lens but very close to the rear face of it.

**4. Acknowledgements**

I am grateful to Graeme Milton and Emmanuel Rashba for reading the manuscript and important comments. I appreciate multiple discussions with Serge Luryi and Michael Shur during my sabbatical stay at the Stony Brook University and during the FTM Workshop of 2006.